\documentclass[12pt,a4paper,final]{iopart}

\usepackage{iopams}  
\usepackage{graphicx}
\usepackage[breaklinks=true,colorlinks=true,linkcolor=blue,urlcolor=blue,citecolor=blue]{hyperref}

\begin{document}

\title{Structural transition in AuAgTe$_4$ under pressure}

\author{A V Ushakov$^{1}$}
\address{$^1$M.N. Miheev Institute of Metal Physics of Ural Branch of Russian Academy of
Sciences, 620108, Ekaterinburg, Russia}
\ead{ushakov@imp.uran.ru}

\author{S V Streltsov}
\address{$^1$M.N. Miheev Institute of Metal Physics of Ural Branch of Russian Academy of
Sciences, 620108, Ekaterinburg, Russia}
\address{$^2$Ural Federal University, Mira St. 19, 620002 Ekaterinburg, Russia}

\author{D I Khomskii}
\address{$^3$ II. Physikalisches Institut, Universitaet zu Koeln, Zuelpicher str. 77, 50937, Koeln, Germany}

\begin{abstract}
Gold is inert and forms very few compounds. One of the most interesting of those is calaverite AuTe$_2$, which has incommensurate structure and which becomes superconducting when doped or under pressure. There exist a ``sibling'' of AuTe$_2$, the mineral sylvanite AuAgTe$_4$, which properties are almost unknown. In sylvanite  Au and Ag ions are ordered in stripes, and Te$_6$ octahedra around metals are distorted in such a way that Ag becomes linearly coordinated, what is typical for Ag$^{1+}$, whereas Au is square coordinated -- it is typical for $d^8$ configurations, i.e. one can assign to Au the valence $3+$. Our theoretical study shows that at pressure P$_\mathrm{C}$ $\sim 5$ GPa there should occur in it a structural transition such that above this critical pressure Te$_6$ octahedra around Au and Ag become regular and practically identical. Simultaneously Te--Te dimers, existing at P $= 0$ GPa, disappear, and material  from a bad metal becomes a usual metal with predominantly Te $5p$ states at the Fermi energy. We expect that, similar to AuTe$_2$, AuAgTe$_4$ should become superconducting above P$_\mathrm{C}$.
\end{abstract}

\vspace{2pc}

\section{\label{sec:Intro}Introduction}
Gold, despite being very inert, can produce solids with quite remarkable properties. One of the most interesting is the materials containing Au and Te -- metallic alloys~\cite{PhysRev.183.619,Meyer-79}, and for certain compositions -- real stoichiometric chemical compounds. Among those the main attention until now was attracted to AuTe$_2$ -- mineral calaverite (we recently predicted that also the compound with 1:1 ratio, AuTe could also exist~\cite{Streltsov2018}). For many years calaverite presented a puzzle for investigators. It is a rare case, of a solid having an incommensurate crystal structure~\cite{VanTriest1990,Ettema1994,Krutzen1999,Ootsuki2014a}. Its puzzling behaviour was finally explained only recently~\cite{Streltsov2018}, as a consequence of spontaneous charge disproportionation in situation with negative charge transfer gap. AuTe$_2$ is also interesting because it is one of very few materials containing Au which become superconducting when doped by Pd or Pt~\cite{Ootsuki2014a,Kudo-13,PhysRevB.5.901}, and also under relative by small pressure of order of 2.6 GPa~\cite{Kitagawa-13}.
\begin{figure}[t!]
\begin{center}
\includegraphics[width=3.5in]{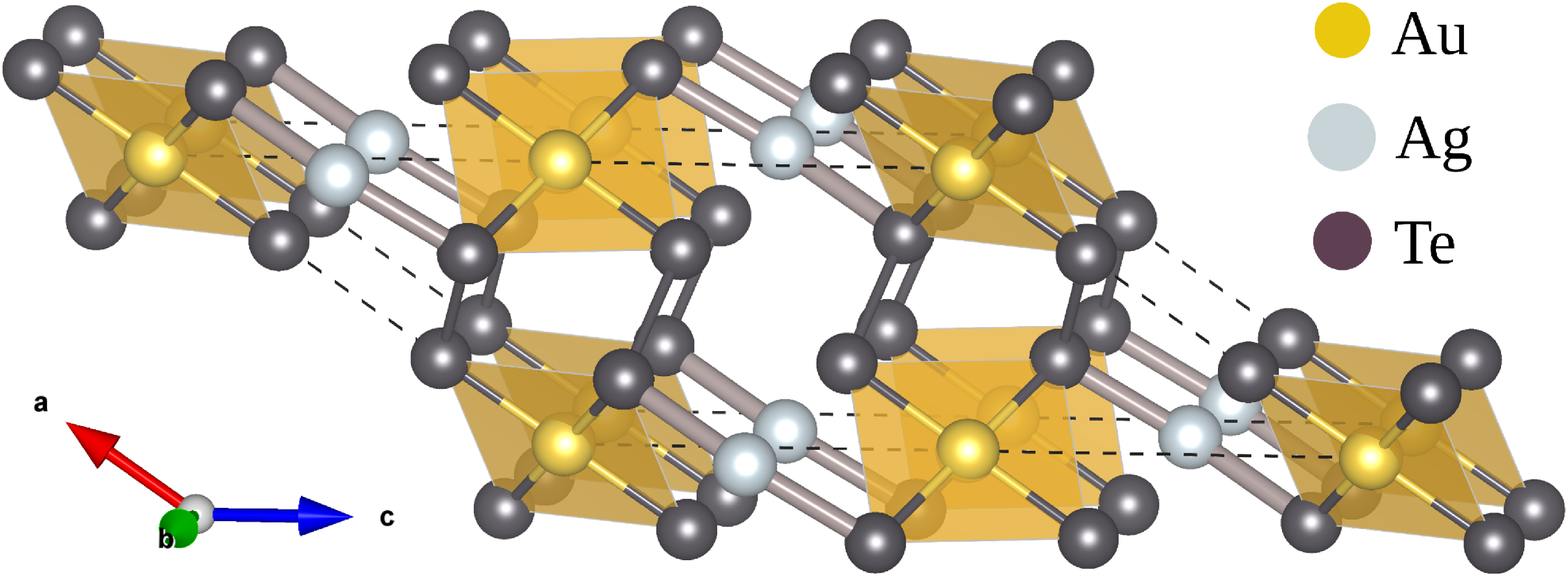}
\includegraphics[width=2.4in]{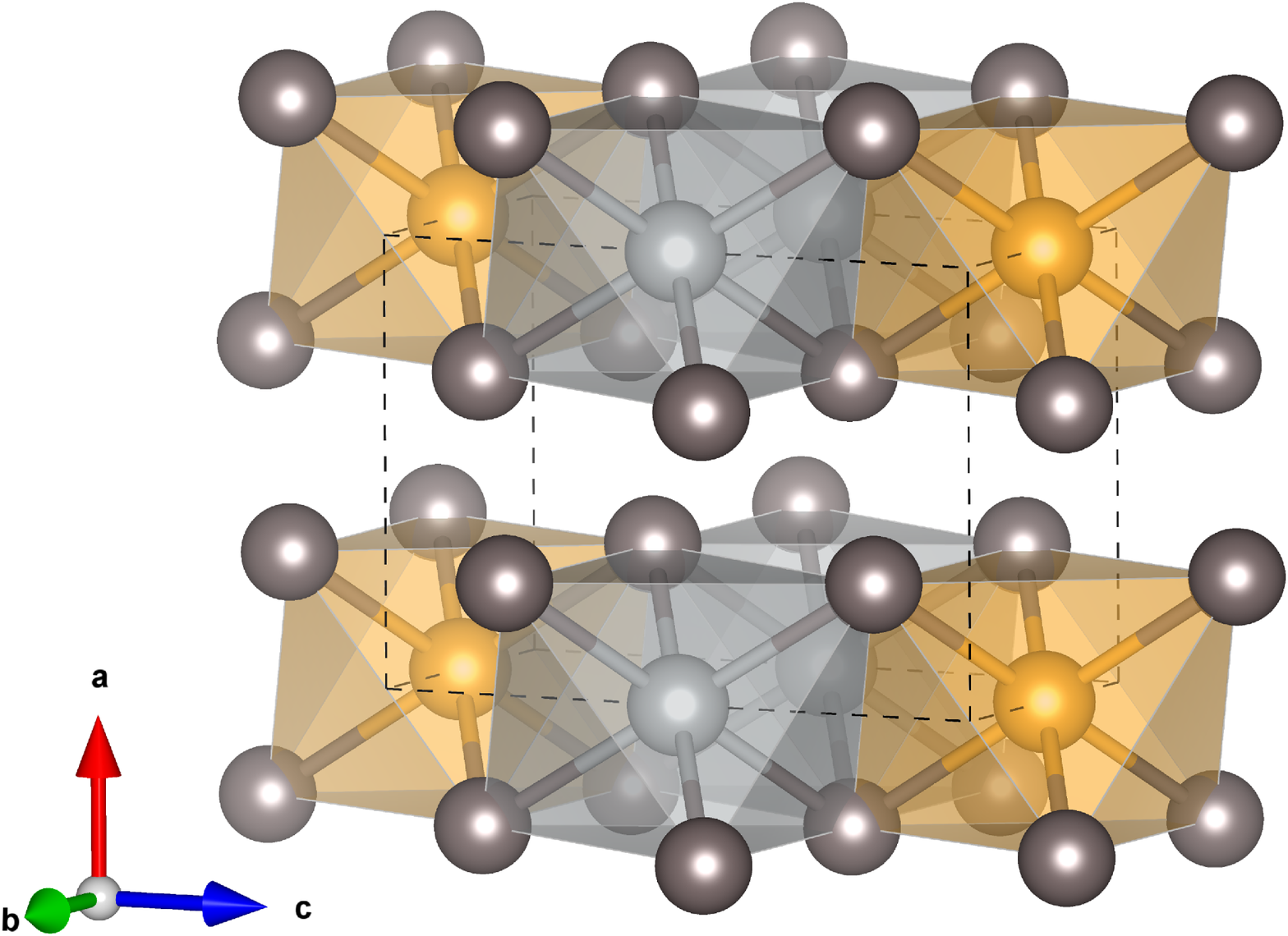}
\end{center}
\caption{\label{struct_ab}(color online) The crystal structure of AuAgTe$_4$ at room temperature and ambient pressure (left figure) and the structure obtained in GGA+SOC approximation at 5 GPa (right figure). The unit cell is marked by a dashed line. The crystal structure was drawn using VESTA~\cite{VESTA} software.}
\end{figure}

There exist in nature related materials, minerals muthmanite AuAgTe$_2$~\cite{Bindi-04,Bindi-08} and sylvanite AuAgTe$_4$. In the sylvanite a half of Au in initial AuTe$_2$ is substituted by Ag, i.e. one has Au$_{\frac{1}{2}}$Ag$_{\frac{1}{2}}$Te$_2$ instead of AuTe$_2$. Sylvanite is characterized by monoclinic space group $P2/c$~\cite{Tunell-52}. In contrast to AuTe$_2$, sylvanite has a regular commensurate ordering of Au and Ag, forming stripes in triangular layers of transition metals (TM), surrounded by Te layers above and below, see Fig.~\ref{struct_ab}. In this sense sylvanite may seem simpler than AuTe$_2$. But it also has interesting features: similar to AuTe$_2$, the average valence of Au and Ag in it is 2+, so that in this sense it may resemble a  pyrite Fe$^{2+}$(S$_2$)$^{2-}$. But both Ag$^{2+}$ and Au$^{2+}$ are unstable and rarely seen in practice, especially Au$^{2+}$; they usually have a tendency to charge disproportionate into 1+ and 3+ ionic states as it occurs e.g. in Cs$_2$Au$_2$Cl$_6$~\cite{Streltsov-11}. Judging from the detailed crystal structure one can conclude that this is indeed what happens in AuAgTe$_4$. Apparently here the valencies of Au and Ag are different. Ag is surrounded by compressed Te$_6$ octahedra, so that it practically becomes linearly coordinated (it has two short bond of $2.73$~\AA, two middle bonds of $2.92$~\AA~and two long bonds of $3.22$~\AA). On the other hand Te$_6$ octahedra around Au are strongly elongated, so that Au have four short bonds of $2.67$~\AA~($2.69$~\AA) and two long bonds of $3.22$~\AA, and is practically square--coordinated, see Fig.~\ref{struct_ab}. This coordination is common for Au$^{2+}$($d^9$) or for Au$^{3+}$($d^8$) states due to a strong Jahn-Teller effect on those (note that, as we just have mentioned, chemically Au$^{2+}$ can hardly be stabilized; and Au$^{3+}$ is a negative charge transfer ion, so that it's real electronic configuration is in fact not $d^8$, but rather $d^9\underline{L}$ or even $d^{10}\underline{L}^2$, where $\underline{L}$ stands for a ligand hole~\cite{PhysRevLett.55.418,Khomskii-97,Sawatzky-16}. Note right away that these distortions also lead to modification of Te sublattice, so that in it short  Te--Te dimers are formed, with two Te's in a dimer belonging to different $M$Te$_2$ planes ($M$ = Ag, Au). Formation of such dimers may provide a rather strong coupling between these layers, so that AuAgTe$_4$ (and similarly AuTe$_2$) should  not be treated as a van der Waals system.

In contrast to a relatively well studied AuTe$_2$, AuAgTe$_4$ attracted much less attention. Encouraged by extraordinary properties of calaverite AuTe$_2$, we undertook a theoretical investigation of sylvanite AuAgTe$_4$, using ab--initio calculations, in particular studying the behaviour of this material under pressure. Quite interestingly, we found that at a pressure of about P$_\mathrm{C}$ $\sim 5$~GPa crystal structure of it should strongly change, so that, first of all, structural distortions disappear and  $M$Te$_6$ octahedra around Au and Ag become regular, with equal Au--Te and Ag--Te distances; and second, despite different chemical elements, these AuTe$_6$ and AgTe$_6$ octahedra become practically identical, with the same $M$--Te bond lengths. Simultaneously with that Te--Te dimers disappear, so that in a sense this material under pressure becomes more two-dimensional. As to electronic structure, at ambient pressure due to Te--Te dimer formation the density of states at the Fermi energy develops a pseudogap, but at P~$>$~P$_\mathrm{C}$ this pseudogap disappears, and this material becomes a regular metal with Te $5p$ states at the Fermi level, so that the system turns out to be a so-called $p$--metal~\cite{Kitagawa-13}. We believe that these changes may lead to the formation of superconductivity at the high-pressure phase of AuAgTe$_4$.

\section{\label{sec:Details}Calculation details}
The electronic structure calculations of AuAgTe$_4$ were carried out using the Vienna Ab-initio Simulation Package (VASP)~\cite{Kresse-93,Kresse-96}. We utilized the projector augmented-wave (PAW) method~\cite{Kresse-99} with the Perdew--Burke--Ernzerhof (PBE) type of exchange-correlation functional within the General Gradient Approximation (GGA)~\cite{Perdew-96}. The energy cutoff was chosen to be $E_{cutoff}=500$~eV and $4\times4\times2$ Monkhorst-Pack grid of k-points was used during the calculations. The crystal structure was relaxed until forces falled behind $0.0005$ eV/\AA. The spin-orbit coupling (SOC) was included to the calculation scheme. The electron population numbers were obtained by integration within
atomic spheres with radii 1.503~\AA, 1.503~\AA~and 1.535~\AA~for Au, Ag and Te correspondingly around each ion.

\begin{figure}[htb!]
\centering
\includegraphics[width=3.5in]{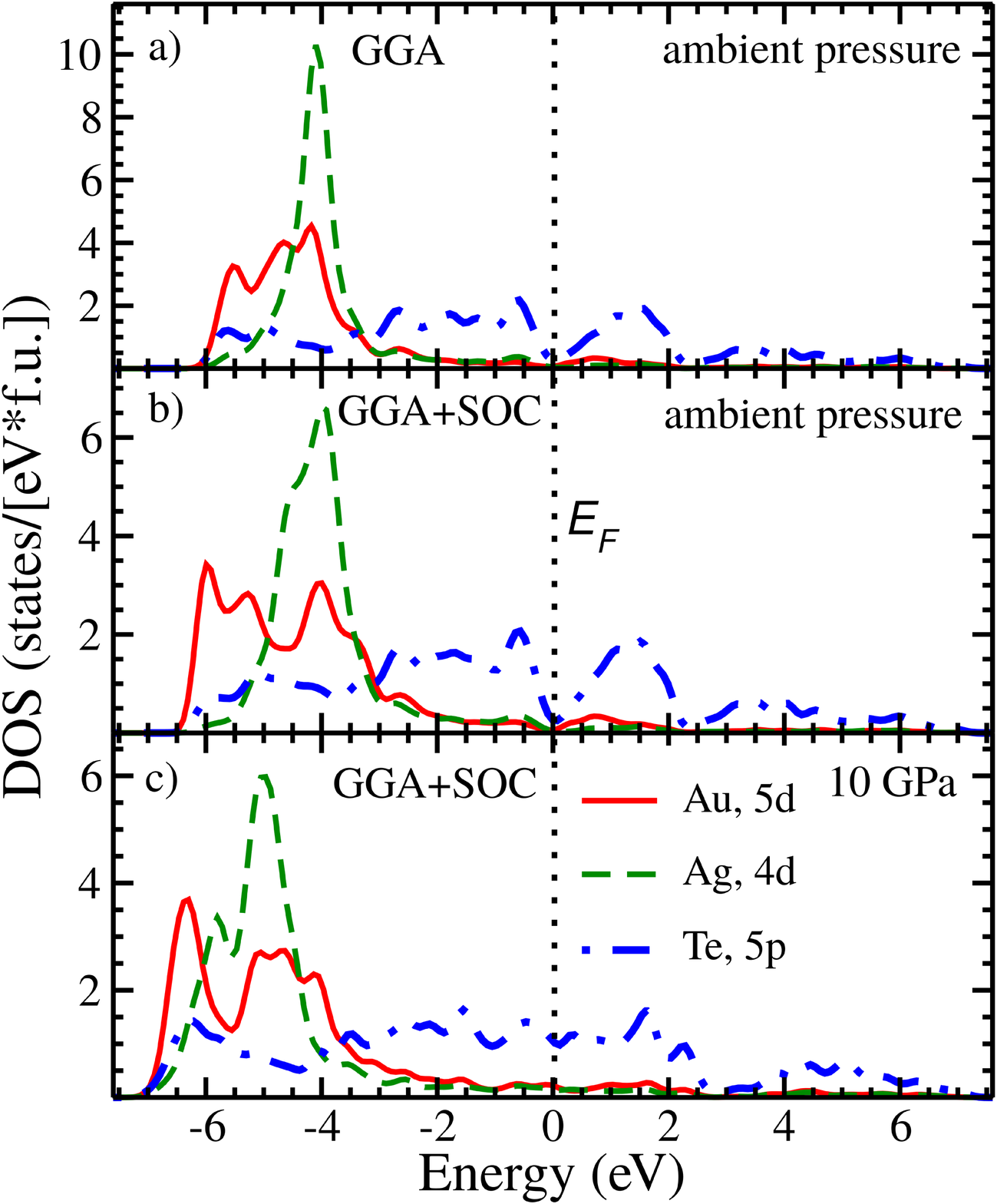}
\caption{\label{struct-dos}(color online) Partial densities of states (DOS)  of AuAgTe$_4$ at ambient pressure and at 10 GPa in the GGA and GGA+SOC approximations. The Fermi energy is at zero.}
\end{figure}

\section{\label{sec:Results} Results }

\begin{figure}[htb!]
\centering
\includegraphics[width=3.5in]{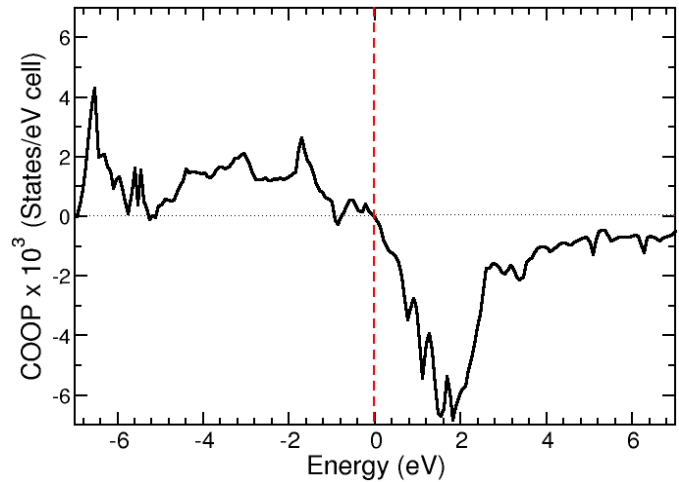}
\caption{\label{coop} The crystal orbital overlap population (COOP) function of AuAgTe$_4$ at 0 GPa. The Fermi energy is at zero.}
\end{figure}

The partial densities of states of AuAgTe$_4$ at normal conditions and at 10 GPa are presented in Fig.~\ref{struct-dos}(a).
At ambient pressure AuAgTe$_4$ has a pseudogap at the Fermi energy, which is due to presence of Te-Te dimers. This can be easily seen from Fig.~\ref{coop}, where the crystal orbital overlap population (COOP) function for Te $5p$ states is plotted (calculated in the local density approximation using the linearized muffin-tin approximation~\cite{PhysRevLett.53.2571}).
The COOP is a very useful tool to study chemical bonding. Positive COOP corresponds to bonding, while negative to antibonding states~\cite{Wills2010}. One may see from Fig.~\ref{coop}, that the Fermi level is almost exactly in the place where the COOP (corresponding to the $p$ states of nearest neighbor Te ions) changes its sign. Thus, the pseudogap in AuAgTe$_4$ appears due to the bonding--antibonding splitting between Te $p$ states in the Te--Te dimer.
\begin{figure}[htb!]
\centering
\includegraphics[width=3.5in]{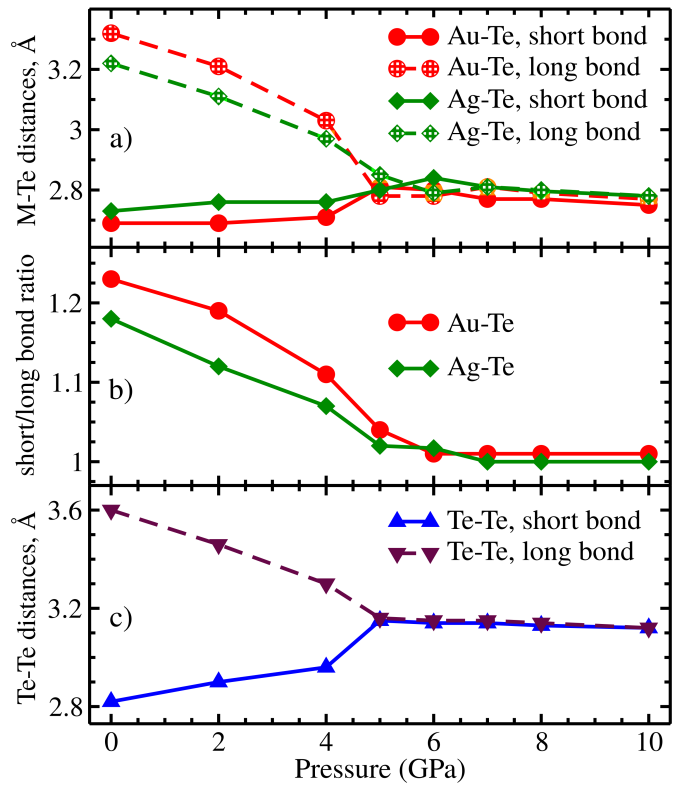}
\caption{\label{struct-evolu}(color online) The evolution of the Au--Te, Ag--Te  ((a) and (b)) and Te--Te (c) bond lengths under the pressure after  optimization of the crystal structure of AuAgTe$_4$ in GGA+SOC approximation.}
\end{figure}

The width of Au $5d$ band is about $4$ eV and it is broader than Ag $4d$ one on $\sim 1$~eV. This is due to larger principal quantum number of covalent $d$ orbitals and plaquette geometry of Au ions. Also the Au $5d$ band lies lower than Ag $4d$ band (on $\sim 0.4$ eV). The spin-orbit coupling shifts positions of both Au $5d$ and Ag $4d$ bands deeper in energy (compare Fig.~\ref{struct-dos}(a) and (b)).

 Another interesting feature of AuAgTe$_4$ electronic structure is that the Te $5p$ band lies higher than Ag $4d$ and Au $5d$ ones. This suggests that AuAgTe$_4$ is also (as AuTe$_2$) in the negative charge transfer energy regime~\cite{Khomskii-97,Sawatzky-16}. It means that the $4d$ and $5d$ holes of transition metal ions will prefer to move to $5p$ shell of Te ions, and Au and Ag ions will have a much larger electronic occupation than one would expect from naive ionic consideration. Indeed there are 9.09 and 9.58 electrons on Au and Ag ions according to our GGA+SOC calculations. It also demonstrates a high level of hybridization between $4d$ ($5d$) TM and Te $5p$ states.

The results of structural optimization of AuAgTe$_4$ under pressure are summarized in Fig.~\ref{struct-evolu} and in Table~\ref{str:tab2}. At 0 GPa the crystal structure of AuAgTe4 is stable, the deviation of the calculated structure from the experimental one is negligible.
\begin{table}[htb!]
\caption{\label{str:tab2} The crystal structure of AuAgTe$_4$ at 5 GPa and at 10 GPa obtained from lattice optimization in the GGA+SOC approximation (see Fig.~\ref{struct_ab}). The space group above 5 GPa was found to be P$2$/$m$.}
\centering
\begin{tabular}{c|ccc|ccc}
\hline
  & \multicolumn{3}{c}{$5$ GPa} &  \multicolumn{3}{|c}{$10$ GPa} \\
\hline
 \multicolumn{7}{c}{crystal structure parameters} \\
\hline
 a, \AA &    \multicolumn{3}{c}{5.1332} &   \multicolumn{3}{|c}{5.0451}  \\
 b, \AA &   \multicolumn{3}{c}{4.1020}  &   \multicolumn{3}{|c}{4.0378} \\
 c. \AA &  \multicolumn{3}{c} {7.1785}  &  \multicolumn{3}{|c}{7.0377}  \\
 $\alpha$ & \multicolumn{3}{c}{90$^{\circ}$} & \multicolumn{3}{|c}{90$^{\circ}$}  \\
 $\beta$ & \multicolumn{3}{c}{90.506$^{\circ}$} & \multicolumn{3}{|c}{90.24$^{\circ}$}  \\
 $\gamma$ & \multicolumn{3}{c}{90$^{\circ}$}  &  \multicolumn{3}{|c}{90$^{\circ}$}    \\
\hline	
 \multicolumn{7}{c}{atomic positions} \\
\hline
 Au,  1b & $0\quad$ & $0.5\quad$ & $0$  &  $0\quad$ & $0.5\quad$ & $0$ \\
 Ag,  1c & $0\quad$ & $0\quad$ & $0.5$  &  $0\quad$ & $0\quad$ & $0.5$  \\
 Te1, 2n & $0.29913\quad$ & $0.5\quad$ & $0.33067$  & $0.30033\quad$ & $0.5\quad$ & $0.33379$  \\
 Te2, 2m & $0.28964\quad$ & $0\quad$ & $0.83586$  & $0.29265\quad$ & $0\quad$ & $0.83649$  \\
\hline
\end{tabular}
\end{table} 
But we see that with pressure a gradual decrease of distortions around Au and Ag takes place, and above critical pressure of P$_\mathrm{C}$~$\sim 5$~GPa the $M$Te$_6$ octahedra become practically ideal with all $M$--Te bond lengths equal (see Fig.~\ref{struct_ab}). The elastic tensor was determined by performing six finite distortions of the lattice and deriving the elastic constants from the strain-stress relationship~\cite{Page-02}. Elastic moduli including contributions for distortions with rigid ions and from the ionic relaxations are presented in Table~\ref{elastic}. The positive values of elastic constants confirm the mechanical stability of calculated AuAgTe$_4$ structures at high-pressure phase.
\begin{table}[htb!]
\caption{\label{elastic} Elastic constants (in GPa) of AuAgTe$_4$ at ambient pressure, 5~GPa and 10~GPa, calculated in the GGA+SOC approximation for monoclinic symmetry. }
\centering
\begin{tabular}{c|ccccccccc}
\hline
& $c_{11}$ & $c_{12}$ & $c_{13}$ & $c_{22}$ & $c_{23}$  & $c_{33}$ & $c_{44}$ & $c_{55}$ & $c_{66}$ \\
\hline
0 GPa  & 96.2  & 13.2 & 16.8 & 35.7  & 26.4 & 61.9  & 11.8 & 20.7 & 16.7 \\
5 GPa  & 192.3 & 33.5 & 36.9 & 123.6 & 90.9 & 117.3 & 13.2 & 60.3 & 16.6 \\
10 GPa & 235.8 & 46.2 & 44.7 & 140.9 & 116.4 &  139.9 & 13.4 & 75.3 & 15.6 \\
\hline
\end{tabular}
\end{table} 

More surprising, structural difference between Au and Ag is lost: all $M$--Te bonds at 10 GPa (above the transition) are $\sim 2.77$ \AA~(the average $M$--Te bond length differs by $0.01$ \AA~in AgTe$_6$ and AuTe$_6$ octahedra). Nevertheless of course these remain different elements. The question is what could be the reason of this surprising behaviour.
The electron occupation of Au $5d$ and Ag $4d$ states at 10~GPa are almost the same as at normal conditions ($9.11$ and $9.54$ electrons for Au and Ag). Thus the charge disproportionation in AuAgTe$_4$ does not disappear under pressure, and Ag and Au ions do not show the electron equivalency in the equivalent surrounding of Te ions.

Some clue can be found in the behaviour of the electronic density of states $\rho(\varepsilon)$ with pressure, especially around the Fermi level. First of all, note that the main contribution to $\rho(\varepsilon)$ close to $E_F$ is provided by the Te $5p$ states, contribution of $d$ states of Au and Ag is being quite small, see Fig.~\ref{struct-dos}. As we stressed above, at ambient pressure due to formation of Te--Te dimers there appears the dip, pseudogap at the Fermi energy, which makes AuAgTe$_4$ a bad metal. And indeed in this phase the usual notions of valence, in particular the usual rules of solid state chemistry connecting valence and electronic configuration of an ion with the structure of its surrounding work quite well (in our case it is linear coordination around TM, here Ag$^{1+}$, with $d^{10}$ configuration, and square coordination for $d^8$ configuration of Au$^{3+}$).

Above P$_\mathrm{C}$, however, the crystal structure changes in such a way that the bond lengths $M$--Te become equal. Simultaneously Te--Te dimerization disappears, see Fig.~\ref{struct-evolu}. In effect the electronic density of states changes significantly: the pseudogap at E$_F$ vanishes (at 10 GPa $\rho({\varepsilon_F})$~$\sim 3$ states/(eV*f.u.), and electronically AuAgTe$_4$ becomes similar to a regular metal. Here the main electron contribution close to the Fermi energy is provided by the Te $5p$ states. One can think that it is just the crossover to a regular metallic state which invalidates the usual notions applicable for localized electrons, such as the effectiveness of Jahn--Teller effect etc. Note in this respect the old idea of John B. Goodenough that there exist two thermodynamically different states of electrons in matter: localized electrons, which in particular can make ions with orbitally--degenerate configurations Jahn--Teller active, and itinerant state, in which Jahn--Teller effect does not work (simply the conditions for its applicability -- the presence of localized electrons, are not satisfied). We can think that the situation in AuAgTe$_4$ above this critical pressure is just that: the material becomes more similar to a regular metal, or rather to Au--Ag--Te alloy, in which Au and Ag $d$--bands lie relatively deep under the Fermi level and lose their localized character. Apparently the situation in calaverite AuTe$_2$~\cite{Kitagawa-13} above critical pressure may be described by the same picture.

One extra conclusion which we can draw from the obtained results and from this picture is that, similarly to AuTe$_2$ at P~$>$~P$_\mathrm{C}$, also AuAgTe$_4$ at the high-pressure phase may become superconducting. Indeed, first of all, it becomes more two dimensional, which may help superconductivity. Second, it apparently becomes a good metal, and with Ag and Au ions becoming structurally identical, they would not induce strong scattering. Such regular metals or metallic alloys may indeed become superconducting if there appears an effective electron--electron attraction leading to Cooper pairing.  For that the specific character of constituting atoms, Au and Ag, may be instrumental. In~\cite{Streltsov2018} we put forth some arguments that just the very well known tendency of Au (and Ag) to charge disproportionation of nominally Au$^{2+}$($d^9$) into Au$^{1+}$($d^{10}$) and Au$^{3+}$($d^8$) (or rather Au$^{1+}$($d^{10}$)$\underline{L} \rightarrow$ Au$^{1+}$($d^{10}$) + Au$^{1+}$($d^{10}$)$\underline{L}^2$) can help superconductivity: this tendency actually means that there exist an ``atomic'' tendency to form electron pairs (it is better to have not one $d$ hole as in Au$^{2+}$($d^9$) but either no holes ($d^{10}$) or two holes ($d^8$ or $d^{10}\underline{L}^2$). I.e. we can say that there acts in such valence skippers something like an effective electron attraction -- effective negative $U$ situation, using the terminology of the Hubbard model. We think that this mechanism can work in favour of making high-pressure phase of AuAgTe$_4$ superconducting. 

\section{Conclusion}

Summarizing, we theoretically obtained that the AuAgTe$_4$, the mineral sylvanite, may strongly change its properties under pressure, from the bad metal with rather strongly distorted lattice, to the state similar to a regular metal, with much less distortions. We presented some arguments that this high-pressure phase of AuAgTe$_4$ may become superconducting. It might be very interesting to try to experimentally check this prediction, all the more so because the critical pressure needed for that is relatively low,  of order of $\sim 5$ GPa.

\section{Acknowledgments}
 Present work was supported by the project of the Ural branch of RAS 18-10-2-37, by the FASO through research programs ``spin'' AAAA-A18-118020290104-2, by Russian ministry of science via contract 02.A03.21.0006 and by the Russian foundation for basic research (RFBR) via grant RFBR 16-32-60070.
 The work of D.I. Khomskii was funded by the Deutche Forschungsgemeinschaft (DFG, German Reseach Foundation), Project number 277146847 -- CRC 1238.

\end{document}